\documentclass[10pt,twocolumn,oneside]{article}
\usepackage[a4paper, total={7in, 9in}]{geometry}
\usepackage[utf8]{inputenc}
\usepackage[T1]{fontenc}
\usepackage{titling}
\usepackage{graphicx}
\usepackage{authblk}
\usepackage{titlesec}
\usepackage{siunitx}
\usepackage{cite}
\usepackage{caption}
\usepackage{fancyhdr}
\usepackage[hidelinks]{hyperref}

\hypersetup{pdfauthor={Jeandet et al.},
            pdfsubject={},
            pdftitle={Spatio-temporal structure of a Petawatt femtosecond laser beam},
            pdfkeywords={Ultrafast lasers, Ultrafast Metrology, High Power Lasers, Ultrashort Pulses}}

\title{\bfseries \Large \vspace{-2cm}Spatio-temporal structure of a Petawatt femtosecond laser beam}
\date{}

\author[1,3]{Antoine Jeandet}
\author[1]{Antonin Borot}
\author[2]{Kei Nakamura}
\author[1]{Spencer W. Jolly}
\author[2]{Anthony J. Gonsalves}
\author[2]{Csaba Tóth}
\author[2]{Hann-Shin Mao}
\author[2,*]{Wim P. Leemans}
\author[1,*]{Fabien Quéré}

\affil[1]{\emph{LIDYL, CEA, CNRS, Université Paris-Saclay, CEA Saclay, 91191 Gif-sur-Yvette, France}}
\affil[2]{\emph{Lawrence Berkeley National Laboratory, Berkeley, CA 94720, USA}}
\affil[3]{\emph{Amplitude Laser Group, Business Unit Science, 2/4 rue du Bois Chaland, 91090 Lisses, France}}
\affil[*]{Corresponding authors: \texttt{fabien.quere@cea.fr}, \texttt{wpleemans@lbl.gov}}

\titleformat{\section}
  {\normalfont\bfseries}{\thesection}{1em}{}

\newcommand \myabstract[2][.8]{%
  \renewcommand\maketitlehookd{%
    \mbox{}\medskip\par
    \centering
    \begin{minipage}{#1\textwidth}
      #2
    \end{minipage}}}

\newcommand\ct[1]{\text{\rmfamily\upshape #1}}

\begin{document}

\myabstract{\bfseries The development of optical metrology suited to ultrafast lasers has played a key role in the progress of these light sources in the last few decades. Measurement techniques providing the complete $E$-field of ultrashort laser beams in both time and space are now being developed. Yet, they had so far not been applied to the most powerful ultrashort lasers, which reach the PetaWatt range by pushing the Chirped Pulse Amplification scheme to its present technical limits. This situation left doubts on their actual performance, and in particular on the peak intensity they can reach at focus. In this article we present the first complete spatio-temporal characterization of a PetaWatt femtosecond laser operating at full intensity, the BELLA laser, using two recently-developed independent measurement techniques. Our results demonstrate that, with adequate optimization, the CPA technique is still suitable at these extreme scales, i.e., it is not inherently limited by spatio-temporal couplings. We also show how these measurements provide unprecedented insight into the physics and operation regime of such laser systems.\\}
\maketitle

\section*{Introduction}
    The technology of ultrafast laser sources nowadays makes it possible to generate laser pulses of femtosecond duration with peak powers of up to several PetaWatts (\si{\peta\watt})~\cite{nakamura17,sung17,zeng17}. The development of such laser sources, now available or under construction in several laboratories worldwide, is motivated by two main prospects. The first is demonstrating compact particle accelerators for scientific or societal applications, with particle energies up to several \si{\giga\electronvolt}~\cite{clayton10,wang13,leemans14,kim17,gonsalves_petawatt_2019}. The second is exploring the physics of ultra-relativistic laser-matter interactions, and more particularly accessing regimes where highly non-linear Quantum Electro-Dynamical effects come into play, in order to perform new tests of this fundamental theory~\cite{marklund06}. 

    Due to their ultrashort durations, such lasers necessarily have broad spectra, typically covering several tens of nanometers in wavelength. As in any optical system using broadband light, chromaticity---i.e., the spectral dependance of the spatial properties of a beam, or of the spatial response of an optical system---can become a major impediment to optimize their performance. PetaWatt ultrashort lasers are particularly exposed to this issue; the key enabling technology for these systems, Chirped Pulse Amplification (CPA), which is pushed to its present technical limits in these systems relies on highly chromatic optical elements such as gratings or prisms, used to tailor the pulses temporal properties for the amplification process~\cite{strickland85}. The large chromatic effects individually introduced by each of these optical elements should ideally perfectly compensate each other at the system output. But any imperfection or misalignment of these optics results in residual chromaticity, as so does the use of simple chromatic elements such as singlet lenses~\cite{bor_distortion_1989}.

    \begin{figure*}[t!]
    \centering
    \includegraphics{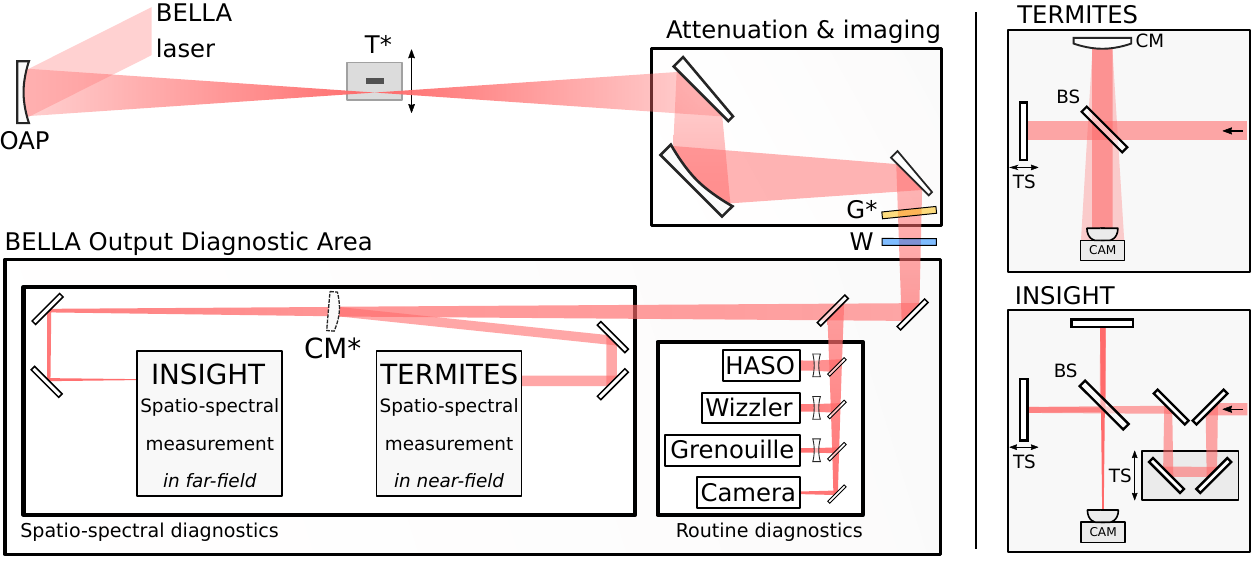}
    \caption{Sketch of the experimental set-up. The left panel shows the path of the BELLA laser beam from the compressor output to the experimental chamber, attenuation and imaging chamber, and finally to the diagnostic area. Here the INSIGHT and TERMITES devices have been installed, together with traditional space-only or time-only diagnostics used for such lasers. The right panels display sketches of the INSIGHT and TERMITES devices, which for practical reasons had to be used successively on two different weeks. The * symbol indicates retractable elements. OAP=off-axis parabola, T=target, G= gold foil, W= Vacuum window, CM= convex mirror, BS= beam splitter, TS=translation stage.}
    \label{fig:1_setup}
    \end{figure*}

    In standard optical assemblies, such as photographic lenses, chromaticity degrades the system performances by affecting its spatial response. The consequences are actually much more severe for ultrashort lasers, where all frequencies not only need to be focused on the same focal spot, but also with the appropriate relative phases, i.e., those leading to the minimal pulse duration. In this case, undesired chromatic effects not only degrade the concentration of energy in space by inducing imperfect overlap of the different frequencies at focus, they can also increase the pulse duration by reducing the spectral width and degrading the spectral phase locally. The combination of these spatial and temporal effects, known in ultrafast optics as spatio-temporal or spatio-spectral couplings, can lead to very significant reductions in peak intensity even for relatively modest chromatic aberrations---especially as the beam size increases and the pulse duration decreases.

    The first step in addressing this issue is the development of adequate metrology tools, which can measure the complete $E$-field of the laser beam in space-time, $E(x,y,t)$ or space-frequency, $\tilde{E}(x,y,\omega)$. Several suitable techniques have been developed in the last 15\,years~\cite{bowlan_crossed-beam_2006,cousin_three-dimensional_2012,wyatt_sub-10_2006,gabolde_single-shot_2006,dorrer_spatio-spectral_2018,miranda14}, and are now becoming more and more broadly used, especially on ultrashort lasers of moderate power. Only recently, two different and complementary techniques, TERMITES~\cite{pariente16} and INSIGHT~\cite{borot18}, have been successfully demonstrated for the characterization of laser systems of up to \SI{100}{\tera\watt} peak power, revealing for the first time the type of spatio-temporal couplings that can affect such complex systems. Yet, none of the existing PW-class lasers have ever been characterized in space-time, thus leaving serious doubts on the actual performance of these cutting-edge sources.

    In this article, we address this key issue by presenting the first complete spatio-temporal characterization of a \SI{1.1}{\peta\watt}--\SI{40}{\femto\second} system, using both TERMITES and INSIGHT, carried out on the BELLA \si{\peta\watt} laser at LBNL, operated \textsl{at full power and intensity on target}. We first detail the implementation of these two techniques and summarize the validation tests that have been performed, and then present the main chromatic effects observed on the beam. We quantify the impact of chromaticity on the laser performance, and explain how the measurements provide insight into the laser system operating conditions and the physics of the beam amplification. 

\section*{Experimental set-up and measurement procedure}

    The BELLA laser, acquired from Thales and operated by LBNL for the last 5\,years, is based on a double-CPA architecture: broadband pulses originating from an oscillator are amplified in a succession of six Ti:Sapphire amplifiers, and a cross-polarized wave contrast enhancement system is installed in between these two CPA systems. The exact system design is detailed in Ref.~\cite{nakamura17}. 

    After temporal compression in the grating compressor, the beam of 20 cm diameter propagates under vacuum and is focused into the main experimental chamber by an off-axis parabola with a focal length $f=\SI{13.5}{\meter}$, producing a focal spot with $w_0 \approx \SI{53}{\micro\meter}$ (half width at $1/\ct{e}^2$ of the field intensity). This chamber is generally used for laser-wakefield particle acceleration experiments, with a capillary discharge plasma as a target (T in Figure~\ref{fig:1_setup}). Here this target was removed from the beam path, letting the laser diverge away from the focus, until it gets large enough to avoid damaging subsequent optics. 

    An all-reflective chromatic-aberration-free telescope is then used to image the best focus, located in the target chamber, to an output diagnostic area, with a magnification factor of 1. The set of relay optics in this telescope includes three uncoated wedges (two in vacuum, one in air) to attenuate laser pulse energy by a factor of $2.7 \times 10^{4}$. When BELLA is operated above \SI{5}{\joule} pulse energy, a thin (\SI{5}{\micro\meter}) self-standing gold-coated foil (G in Figure~\ref{fig:1_setup}) is inserted into the beam for a further attenuation by a factor 40. This unique attenuation system is what allows the beam to be safely operated at full power inside the vacuum chamber, and therefore also at full intensity at focus, while avoiding optical damage or nonlinear effects in the subsequent diagnostic line (in particular in the exit vacuum window W in Figure~\ref{fig:1_setup}).

    The diagnostic area located after this attenuation system incorporates the spatial-only and temporal-only measurement devices typically used for such lasers, as indicated in Figure~\ref{fig:1_setup}. In the present work, we have implemented two additional measurement devices, based on the TERMITES and INSIGHT techniques, in order to determine the spatio-temporal field of the BELLA laser beam. 

    TERMITES is typically implemented on collimated beams. We therefore recollimated the converging BELLA image beam to a diameter of about \SI{1.8}{\centi\meter}, by inserting a convex spherical mirror prior to the focus image plane (CM in Figure~\ref{fig:1_setup}). TERMITES then uses a Michelson interferometer with a flat mirror in one arm, and a convex spherical mirror in the other. In this second arm, the central part of the beam is converted into a diverging beam, which eventually fully overlaps with the beam to be characterized and is used as a reference, as in point diffraction interferometry. The spatio-spectral comparison of these two beams is achieved by measuring, on each pixel of the detector, the signal resulting from their superposition as a function of delay. Their cross-spectral density function is then obtained by a Fourier transformation of this first order cross-correlation function. 

    INSIGHT also relies on a Michelson interferometer, now with a flat mirror in each arm. This is used to measure the spatially-resolved first order autocorrelation function of the beam at each point of the detector (standard Fourier-transform spectroscopy), from which one can deduce the frequency-resolved spatial intensity profile of the beam. This measurement is performed for multiple planes at and around best focus (here, around the image plane of the best focus). These data are then injected into an iterative phase-retrieval algorithm of the Gerchberg-Saxton type, to determine the spatial phase profile of the beam \textit{at each frequency}.

    \begin{figure}[t!]
    \centering
    \includegraphics{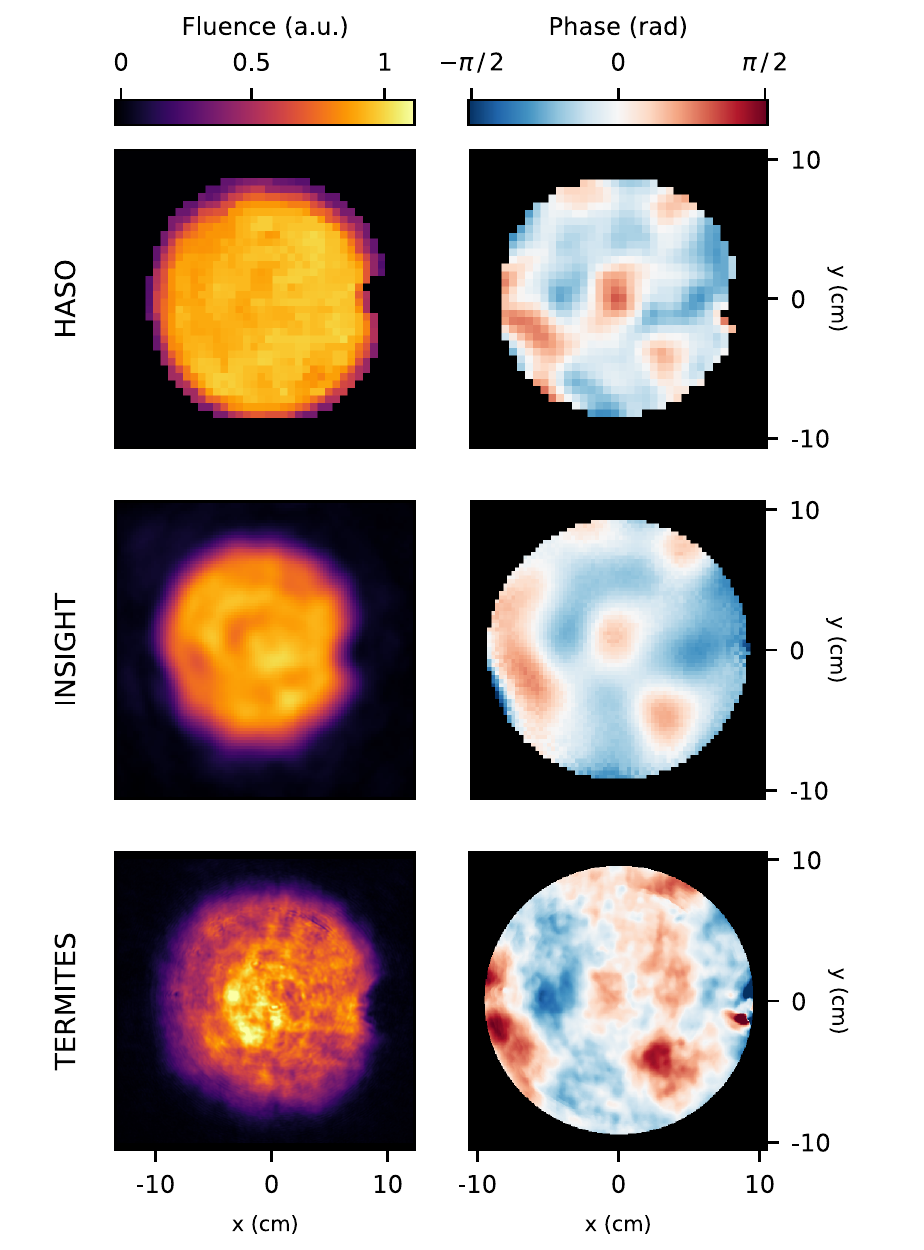}
    \caption{Validation test of TERMITES and INSIGHT on the BELLA laser. Results derived from these two techniques (middle and bottom rows, obtained after frequency-averaging of the spectrally-resolved measurement outputs) are compared to the ones provided by a standard Shack-Hartmann wavefront sensor (HASO by Imagine Optics, upper row). The left column shows the frequency-averaged spatial intensity profile of the collimated BELLA beam, and the right column its frequency-averaged spatial phase profile. Note the small clip on the right edge of the beam: this is due to a small hole in the first wedge of the attenuation line, normally used to collect the accelerated electron beam in laser-wakefield experiments. This hole is usually placed at the center of the laser beam, but was moved to its edge for this measurement campaign. The resulting clip is well retrieved by TERMITES and INSIGHT.}
    \label{fig:2_valid}
    \end{figure}

    Once the spatial amplitude and phase profiles at each frequency have been determined in a given plane (different for TERMITES and INSIGHT), the beam can be numerically propagated to any arbitrary longitudinal position, using plane wave decomposition~\cite{goodman_introduction_2005} at each wavelength. All results presented in this article correspond to the front focal plane of the focusing parabola, located at a distance $f=\SI{13.5}{\meter}$ \textsl{before} this optic.

    Like most of their relatives, these two techniques should actually be considered as spatio-spectral techniques, which determine $\tilde{E}(x,y,\omega)=\tilde{A}(x,y,\omega)\exp\left[i \phi(x,y,\omega)\right]$ up to an unknown spatially-homogeneous spectral phase. An additional measurement of the spectral phase at a single point of the beam is then required to lift this unknown, after which the spatio-temporal field $E(x,y,t)$ can be calculated by a Fourier-transform with respect to frequency. In the present case, this complementary information was provided by the GRENOUILLE device located in the output diagnostic area.

    Both techniques require from a few hundreds to a thousand laser shots for each measurement, and all measurements were performed with the laser operated at \SI{1}{\hertz}. 
    Three operation modes of the laser were used: 'low' power mode (\SI{225}{\milli\joule} per pulse, \SI{5.8}{\tera\watt} peak power) with the last amplifier turned off and wedge attenuation in the chain, moderate power mode (\SI{8.4}{\joule} per pulse, \SI{215}{\tera\watt} peak power) in the same conditions but without attenuation, and full power (\SI{43}{\joule} per pulse, \SI{1.1}{\peta\watt} peak power) with all amplifiers on and no attenuation before target.

    \begin{figure*}[t!]
    \centering
    \includegraphics{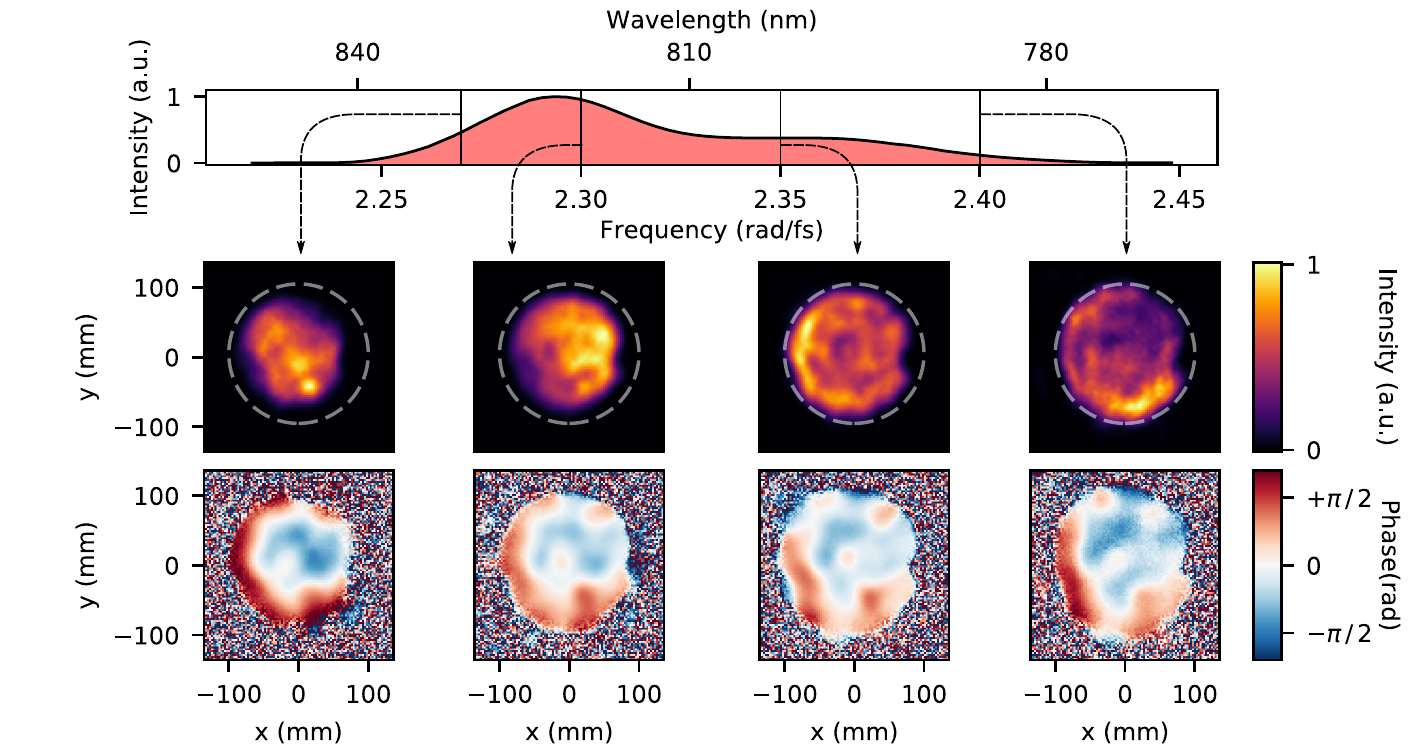}
    \caption{Frequency-resolved spatial properties of the collimated BELLA beam. Four frequencies $\omega_i$ are selected in the spatially-integrated spectrum (upper panel) and the corresponding spatial intensities $\tilde{A}^2(x,y,\omega_i)$ and phases $\phi(x,y,\omega_i)$ (lower row) are shown. These results have been obtained with INSIGHT.}
    \label{fig:3_main-results}
    \end{figure*}

    We performed a set of tests with both techniques to verify that the attenuation and transport system does not lead to spatio-temporal distortions of the laser beam. First, to test the potential influence of non-linear optical effects occurring along this system, we checked that the results of measurements performed at full power with Fourier-transform limited (FTL) pulses were identical to those obtained with pulses of the same energy, but chirped in time to \SI{1.2}{\pico\second} by moving one of the compressor gratings (see Supplemental Document). For the consistency of this test, we also verified that chirping the pulse does not modify the spatio-temporal structure of the beam (due e.g., to undesired angular motion of the compressor grating), by comparing measurement results obtained at low power (thus safely avoiding any non-linear effect in the attenuation system) for FTL and chirped pulses. Finally, at moderate power, we compared the results of measurements carried out with and without the gold foil, which showed that this foil does not introduce any additional coupling on the transmitted beam, but only a smooth spatially-homogeneous modification of the spectral envelop, corresponding to its theoretical transmission curve. With the support of all these tests, all results presented in this article correspond to the optimally-compressed BELLA beam at full power.

    Validation tests of the measurement techniques themselves have already been presented, in Ref.~\cite{pariente16} and~\cite{borot18} for TERMITES and INSIGHT respectively. The measurement configuration used in the present work offered an opportunity for a new type of test, which had not been possible before for practical reasons: with the measurement devices separated from the Shack-Hartmann wavefront sensor (HASO by Imagine Optic) of the diagnostic area by only a few plane mirrors of small diameter, we can safely compare the spatial intensity and phase profiles provided by this well-established instrument, to the information provided by TERMITES and INSIGHT. Since this sensor has no spectral resolution, this is achieved by frequency-averaging the spectrally-resolved spatial profiles determined by these new techniques. This comparison, presented in Figure~\ref{fig:2_valid}, reveals a very good agreement, especially for the phase profiles, which is quite remarkable given the very different principles underlying these three techniques. 

\section*{Results and discussion}

    Spectrally-resolved spatial intensity and phase profiles of the BELLA laser at four different frequencies, measured at full power with INSIGHT, are displayed in Figure~\ref{fig:3_main-results}. The complete evolution of these profiles as a function of frequency is displayed in Supplementary Movie 1.

    The first important conclusion is that BELLA does not suffer from major low-order chromatic phase aberration. Such aberrations can be particularly detrimental for ultrashort lasers, and are expected to be of two main types for collimated laser beams: (i) angular dispersion, typically resulting from stretcher or compressor misalignment, associated in the time domain to pulse front tilt (PFT), (ii) frequency-dependent wavefront curvature,  typically induced by chromatic lenses, associated in the time domain to pulse front curvature (PFC). 

\begin{figure}[t!]
    \centering
    \includegraphics{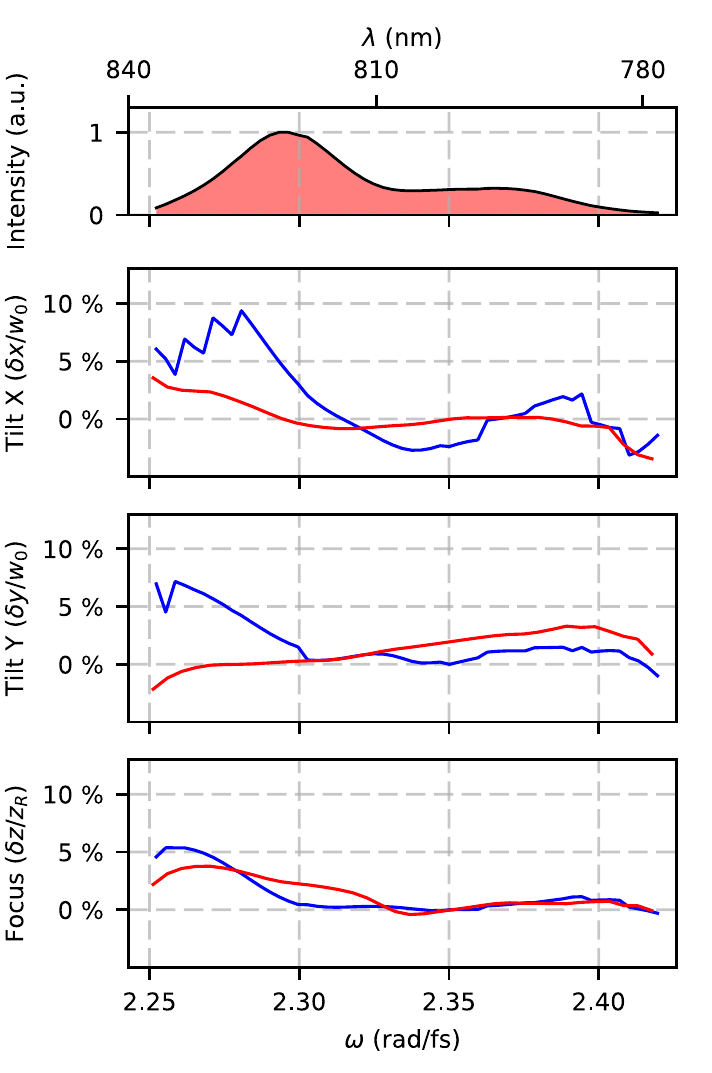}
    \caption{Low-order coefficients (tilt and curvature) of the Zernike polynomials decomposition of the spectrally-resolved spatial phase retrieved with TERMITES (red) and INSIGHT (blue). Tilt values are converted to lateral displacement of the focal spot, relative to the beam waist. Equivalently, curvature values are expressed in longitudinal displacement of the focal spot, relative to the Rayleigh range of the focused beam.}
    \label{fig:4_zernike-modes}
    \end{figure}

    The magnitude of these couplings is observed to be very weak on BELLA, as clearly appears by considering the coefficients of the frequency-resolved Zernike decomposition of the spatial phase, displayed in Figure~\ref{fig:4_zernike-modes}. At focus, the variation of the lateral position of the focal spot, due to frequency-dependent horizontal and vertical wavefront tilts, is less than \SI{10}{\percent} to \SI{15}{\percent} of the beam waist across the laser spectrum. The wavefront curvature does exhibit a variation with frequency, slightly increasing on the red side of the spectrum (see also Figure\ref{fig:3_main-results}). Yet, at focus, this curvature results in a longitudinal focus shift that never exceeds 8\% of the BELLA Rayleigh range. This variation is not linear with frequency, indicating a higher-order spatio-spectral aberration, as opposed to a simple PFC.

\begin{figure}[t!]
    \centering
    \includegraphics{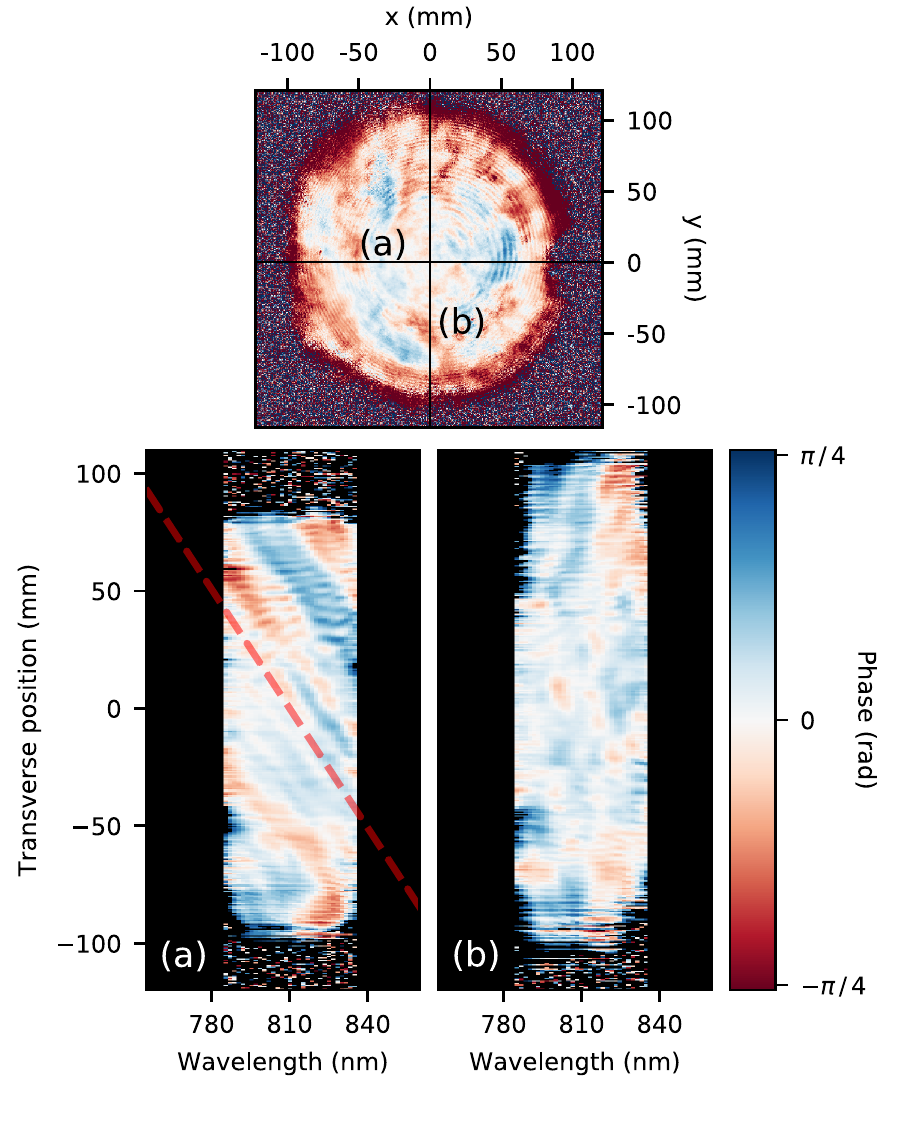}
    \caption{Spatio-spectral phase of the collimated BELLA beam, measured with TERMITES. Plots (a) and (b) respectively display $(x,\omega)$  and $(y,\omega)$ slices of this spatio-spectral phase, across the horizontal  and vertical planes, at the center of the beam (see lines in the upper panel showing a 2D phase profile at one frequency, with the spectrally-averaged wavefront subtracted). The red dashed line in (a) shows the estimated slope of the oblique spatio-spectral pattern that would be induced by spatial defects of optical elements located within the compressor, where the beam is laterally chirped.}
    \label{fig:5_sliding-patterns}
    \end{figure}

    Figure~\ref{fig:4_zernike-modes} also shows that the results obtained with TERMITES for the low-order chromatic phase aberrations are qualitatively consistent with those provided by INSIGHT, especially when considering the fact that these measurements were carried out on different weeks. But TERMITES actually provides finer information on higher-order chromatic aberrations affecting the beam. This clearly appears in Supplementary Movie 2, which shows the complete evolution with frequency of the beam spatial phase, provided by TERMITES: on these phase plots, spatial patterns of higher-frequency are observed, which drift along the horizontal direction as frequency varies. This effect is illustrated more quantitatively in Figure~\ref{fig:5_sliding-patterns}, which displays $(x,\omega)$  and $(y,\omega)$ slices of this spatio-spectral phase, in the horizontal and vertical planes, at the center of the beam ($x=y=0$). The horizontal plane ($x$ axis) corresponds to the dispersion plane of the compressor. At each wavelength, 'high-frequency' (5-6 periods across the full beam) modulations of the spatial phase are observed along $x$. These modulations linearly shift in space as frequency changes, creating an oblique spatio-spectral pattern in the $(x,\omega)$ space (Figure~\ref{fig:5_sliding-patterns}a). By contrast, such an oblique spatio-spectral pattern is not observed at all in the $(y,\omega)$ space, i.e., in the plane normal to the compressor dispersion direction (Figure~\ref{fig:5_sliding-patterns}b).

    A simple interpretation of this feature has been recently analyzed in detail in Ref.~\cite{li17}. In between the first and the last gratings of the compressor the laser beam is laterally chirped, i.e., the central position of the 'beamlet' associated to each frequency shifts approximately linearly with frequency. When this occurs, any modulation of the spatial phase induced on the beam, e.g., due to defects of a mirror or a grating, is imprinted at a different position relative to the center of each beamlet. After the compressor, this lateral chirp is removed, and all beamlets are again centered on the same axis. The spatial modulations induced within the compressor are then expected to shift approximately linearly with frequency, precisely as observed in Figure~\ref{fig:5_sliding-patterns}. The expected slope of the pattern induced in this way can be deduced from the geometry of the compressor and the characteristics of its gratings, and is consistent with the observed effect (red dashed line in Figure~\ref{fig:5_sliding-patterns}), thus strongly supporting this interpretation.

    A simple analytical calculation (see Supplemental Document) shows that, if the corresponding spatial phase modulation is assumed to be perfectly sinusoidal at each frequency and of weak amplitude, then the scattered energy at focus forms two satellite spots located at a distance of $\lambda_0f/\Lambda$ ($\Lambda$ is the spatial period of the modulations) on each side of the main focal spot, with one satellite pulse advanced in time and the other delayed with respect to the main pulse, by $\pm 2\pi s/{\Lambda}$ ($s$ is the slope of the modulation). For the parameters seen in the measurements presented here ($\lambda_0=\SI{810}{\nano\meter}$, $f=\SI{13.5}{\meter}$, $\Lambda\approx\SI{4}{\centi\meter}$, $s=\SI[per-mode=symbol]{0.445}{\meter\femto\second\per\radian}$ or $\SI[per-mode=symbol]{1.31}{\milli\meter\per\nano\meter}$ as in the figure) this results in an offset of \SI{273}{\micro\meter} and \SI{71}{\femto\second} in space and time respectively. With an amplitude of \SI{0.4}{\radian} peak-to-valley over most of the beam, the satellite foci only contain \SI{2}{\percent} of the beam energy, and are at least 100$\times$ less intense than the main pulse.

    Because of the relatively low amplitude of these modulations, most of the associated signal falls within the background of images acquired around focus. As a result, they are not retrieved in the near-field when using the INSIGHT measurements. More generally, TERMITES tends to have a much higher spatial resolution than other techniques for reconstruction of collimated beams: this clearly appears when comparing the different profiles in Figure~\ref{fig:2_valid}. This is because resolving the spatial interference fringes generated by the TERMITES interferometer requires a very high spatial sampling of the collimated beam (above 2000$\times$2000 pixels within the beam profile here), which is preserved in the final reconstruction of the beam.

    To quantify the impact of these various phase aberrations on the laser performance at focus, we define and calculate two different spatio-temporal Strehl ratios of the beam (See the Supplemental Document for details). The first one, $\textrm{SR}_\textrm{Full}$, is the full spatio-temporal Strehl ratio: it is defined as the ratio of the actual peak intensity at focus, derived from the measured $E$-field, to the peak intensity that would be obtained for a beam with the same spatio-spectral amplitude, but with a perfectly constant spatio-spectral phase. This accounts for all phase aberrations present on the beam, both chromatic (i.e., frequency-dependent) and achromatic. We find $\textrm{SR}_\textrm{Full}=0.62$ from TERMITES data, and $\textrm{SR}_\textrm{Full}=0.88$ from the INSIGHT data. We note that the contribution of the oblique spatio-spectral pattern induced by the compressor (detected by TERMITES and not by INSIGHT) to this intensity reduction is actually very weak, so this cannot account for the different values of $\textrm{SR}_\textrm{Full}$ predicted by the two techniques.

    The second Strehl ratio of interest, $\textrm{SR}_\textrm{STC}$, is defined as the ratio of the peak intensity at focus now calculated with the frequency-averaged wavefront distortions removed, to the peak intensity that would be obtained with a perfectly constant spatio-spectral phase. This corresponds to the Strehl ratio that would be achieved if the laser beam wavefront was corrected by a \textit{perfect} adaptive optical system, removing all achromatic phase aberrations---while the deformable mirror actually used in the experiment still leaves some residual frequency-independent wavefront distortions (displayed in Figure \ref{fig:2_valid}). This optimally-corrected beam would then still suffer from chromatic aberrations, which we detect in the present work. We find $\textrm{SR}_\textrm{STC}=0.85$ for TERMITES, and $\textrm{SR}_\textrm{STC}=0.92$ for INSIGHT. This shows that in the case of BELLA, the intensity reduction induced by purely chromatic effects is limited.

    These calculations indicate that the significant difference in values of $\textrm{SR}_\textrm{Full}$ derived from the two techniques, mostly originate from differences in the frequency-averaged beam properties, rather than from chromatic effects. The TERMITES and INSIGHT measurements were conducted one week apart, and indeed measurements with a standard wavefront sensor (HASO) indicate a lower frequency-averaged Strehl ratio in the week of the TERMITES measurements (see Supplemental Document). However, this difference is too weak to fully account for the one found on $\textrm{SR}_\textrm{Full}$between the TERMITES and INSIGHT results. This could be due to differences in the additional aberrations introduced in the transport line or measurement devices themselves, or to a more accurate calculation of the Strehl ratios from the TERMITES data due to the very high spatial resolution of this technique. More systematic and thorough measurement campaigns on smaller laser systems will be required to elucidate this point. In any case, along with the introduction of these calculations, the relatively good agreement and low severity of the $\textrm{SR}_\textrm{STC}$ values are key results of this work.

    \begin{figure}[t!]
    \centering
    \includegraphics{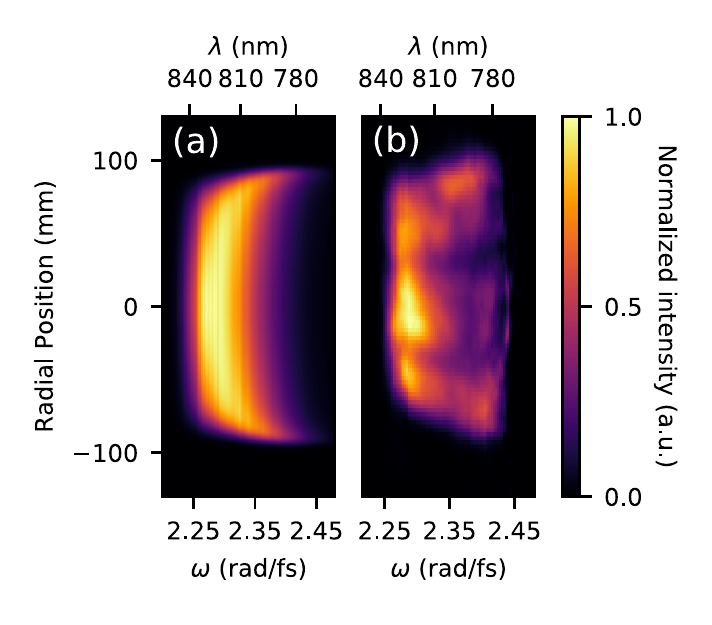}
    \caption{Spatio-spectral intensity of the BELLA beam. The left image displays the $(x,\omega)$ slice of this spatio-spectral intensity, across the horizontal plane, at the center of the beam ($x=y=0$), deduced from simulations of the CPA process based on the Frantz-Nodvik equations. The right image shows the same quantity, now derived from INSIGHT measurements at full power. To clearly highlight the spectral evolution of the beam spatial profile in these plots, the total energy in the beam at each frequency has been normalized to 1  (i.e., $\int dxdy \tilde{A}^2(x,y,\omega)=1$ for all $\omega$ within the laser spectrum).}
    \label{fig:6_beam-size}
    \end{figure}

    We have so far concentrated on the phase properties of the BELLA beam, but our measurements actually also reveal interesting effects on the spatio-spectral intensity: as observed in Figure~\ref{fig:3_main-results}, the beam mode size and shape significantly vary with frequency. As the frequency increases, the beam diameter increases, and the beam profile turns to a ring shape. The fact that the clip imprinted on the right edge of the beam (see Figure~\ref{fig:2_valid} and explanations in its caption) is observed on the blue side of the spectrum but not on the red one clearly indicates that this is not a measurement artifact. We also emphasize that the reconstruction of such fine features in the near-field beam profile from INSIGHT data is quite remarkable, since the measurement was actually performed around focus, and the beam was then numerically propagated to a plane far from focus. This provides a strong validation of the spatial phase retrieved at focus with INSIGHT. 

    We can use simple numerical modeling to show that this frequency-dependent intensity profile is an intrinsic feature of high-power CPA lasers, which had not been clearly identified so far, to the best of our knowledge. To this end, we have modeled the amplification of the  BELLA beam in the successive amplifiers, using the Frantz-Nodvik equations and considering a positively chirped laser beam (i.e., lower frequencies arrive first). Details of these simulations are provided in the Supplemental Document, and essentially start with a measured spectrum after the XPW stage (where the spectral phase is flat), and the spectral phase imparted by the stretcher immediately after. This beam is then amplified in the successive five amplifier stages, with the pump and seed modelled in space, and the full spectral gain cross-section taken in to account. In Figure~\ref{fig:6_beam-size}, the $(x,\omega)$ spatio-spectral intensity profile of the beam predicted by these simulations for full power operating conditions (panel a) is compared to the one deduced from our INSIGHT measurements (panel b, same data as in Figure~\ref{fig:3_main-results}), revealing a good qualitative agreement. 

    These simulations can then be exploited to get a simple interpretation of this effect: in the CPA scheme, the lower frequencies arrive first in the amplification medium, and are more efficiently amplified in the center of the beam. These frequencies deplete the amplification medium before the whole chirped pulse has been amplified. As a result, higher frequencies, which arrive later, are more efficiently amplified on the beam edges than at its center. This qualitatively explains the spectral evolution of the beam size, and the ring shape observed at high frequencies.

\section*{Conclusion}

    In conclusion, we have reported the first spatio-temporal characterization of a PW femtosecond laser, carried out with two independent measurement techniques, with the laser focused at full power in the experimental chamber. These measurements show that the BELLA laser does not suffer from major chromatic aberrations that would severely degrade its performance. We estimate the reduction in peak intensity resulting merely from spatio-temporal couplings to about \SI{10}{\percent}, thus demonstrating that the CPA technique is still suitable and manageable at these extreme scales. We note however that significant chromatic effects start appearing on the red edge of the spectrum, suggesting that the beam optimization would possibly get more challenging for lasers with broader spectra, such as PW lasers now under construction in different laboratories worldwide.

    These measurements provide very fine information on the beam properties and the laser system operation. In particular, we have observed small spatio-spectral modulations that have recently been predicted~\cite{li17} to result from defects of optics within the grating compressor, as well as effects resulting from the temporal dynamics of the gain during the amplification of the chirped laser pulses. As ultrahigh intensity laser-plasma interaction experiments put more and more stringent requirements on the control of the laser parameters, such information could be used in the future for extremely fine optimization of high-power laser systems. 

    Laser-matter interaction experiments should ultimately greatly benefit from this new degree of knowledge of ultrashort laser beams. For instance, in the case of BELLA, the full spatio-temporal profile of the laser beam can be calculated in any arbitrary plane (see Supplementary Movie 3), and this can be used as an input for a realistic description of the laser field in 3D Particle-In-Cell simulations. This opens the way to a deeper understanding of laser-wakefield electron acceleration experiments on this system, which require guiding of the beam in plasma channels of many centimeters in length. This could ultimately contribute to an unprecedented optimization of laser-plasma accelerators. 

\section*{Funding Information}
    The research leading to this work has been funded by the ERC (grant ExCoMet number 694596), by Investissements d'Avenir LabEx PALM (projects EXYT and IMAPS), by the CEA's DFR Impulsion program, and has received financial support from Amplitude Laser Group. The work at the BELLA facility was supported by the Director, Office of Science, Office of High Energy Physics of the U.S. Department of Energy under Contract No. DE-AC02-05CH11231.

\section*{Acknowledgements}
    The authors gratefully acknowledge Amplitude Laser Group for their support. We also thank Chris Pieronek for helping with the experiment, Donald Syversrud, Zachary Eisentraut, David Evans, Nathan Ybarrolaza and Greg Mannino for technical support.

\bibliographystyle{iopart-num}
\bibliography{./biblo_BELLA}

\end{document}